\documentclass{sig-alternate-05-2015}

\usepackage{paralist}
\usepackage{xspace}
\usepackage{newclude}
\usepackage[font=bf]{caption}
\usepackage{subcaption}
\usepackage{graphicx}
\usepackage[hidelinks]{hyperref}
\newcommand{\appendixversion}{1}
\begin{document}
\newcommand{\etal}{\textit{et al}.\xspace}
\newcommand{\ie}[1]{\textit{i.e.\ }#1}
\newcommand{\eg}[1]{\textit{e.g.\ }#1}
\newcommand{\vs}{\textit{vs.\ \xspace}}
\newcommand{\todo}[1]{\textsc{#1}}
\setcopyright{rightsretained}





%

\title{Cloud-Based Distributed Mutation Analysis}
%
%
%
%
%

\numberofauthors{2} 
%
\author{
%
%
\alignauthor
Robert Merkel\titlenote{corresponding author}\\
       \affaddr{Faculty of Information Technology}\\
       \affaddr{Monash University}\\
       \affaddr{Clayton, Victoria, Australia}\\
       \email{robert.merkel@monash.edu}
\alignauthor
James Georgeson\\
       \affaddr{Atlassian}\\
       \affaddr{Sydney, NSW, Australia}\\
       \email{jegeorgeson@gmail.com}
}
\maketitle
\begin{abstract}


Mutation Testing is a fault-based software testing technique which is 
too computationally expensive for industrial use. Cloud-based distributed computing 
clusters, taking advantage of the MapReduce programming paradigm, represent 
a method by which the long running time can be reduced.  In this paper, we describe an architecture, and a prototype implementation, of 
such a cloud-based distributed mutation testing system.  To evaluate the system, we compared the performance of the prototype, with various cluster sizes, to an existing 
``state-of-the-art'' non-distributed tool, PiT.  We also analysed different approaches to work
distribution, to determine how to most efficiently divide the mutation analysis task.  Our tool outperformed 
PiT, and analysis of the results showed opportunities for substantial further performance improvement.

\end{abstract}


%



\section{Introduction}
Mutation Testing, also known as Mutation Analysis or Program Mutation, is a fault-based software
    testing technique that can be used to assess the comprehensiveness of a program's test suite.
    Many \emph{mutant} versions of a program are generated, each of which differs from the original
    by the inclusion of some small fault such as the replacement of the statement \texttt{y=x+2}
    with \texttt{y=x-2}. The original program's test suite is then executed against each mutant. If
    the inclusion of the mutant's fault causes some test case to \emph{fail}, then the mutant is
    said to be \emph{killed}.  Test suite quality is then measured
    by calculating the proportion of mutants killed by the test suite, the \emph{mutation score} (or \emph{mutation adequacy score}).
    
    Contemporary software development practices have increased the importance of automated assessments of test suite quality. The practices of continuous integration~\cite{beck1998extreme}, 
    and continuous delivery~\cite{humble2010continuous} mean that test suites are used---and modified---many times throughout a working day.  Furthermore, after testing, software systems are often deployed to production use on multiple occasions per day.  Delivering reliable software
    in these circumstances requires fast, automated, and accurate assessments of test suite quality.  
    
    Despite considerable evidence that mutation analysis scores are a more accurate estimate of a 
    test suite's ability to detect software faults than conventional coverage metrics~\cite{daran_software_1996, andrews_is_2005,li_experimental_2009}, it is yet
    to achieve significant industrial adoption. Jia and Harman~\cite{jia_analysis_2011}
    suggest that the slowness of mutation analysis is a major reason for this. Generating many mutant versions of a program and executing each of them 
    against a test suite is a computationally expensive process---in one study~\cite{andrews_is_2005}, an approximately 6000 line C program 
     produced over 10000 mutants, each of which must be executed against large parts of the software's test suite.

    In other sectors of information technology, cloud computing has allowed businesses to utilise
    technology in ways not previously possible. Services such as
    Amazon's EC2~\cite{amazon_web_services_amazon_2015} provide users with elastic access to
    computational resources as needed without requiring a significant upfront hardware
    investment~\cite{marston_cloud_2011}. A
    mutation testing system designed for cloud architectures could provide software developers with
    results in acceptable timeframes, making it viable even for large industrial projects.
    
    In this paper, we describe an architecture for a cloud-based, distributed mutation analysis tool, and present a prototype tool based on this architecture.  Our tool
    makes use of Apache Spark~\cite{zaharia_spark:_2010}, a framework for distributed cloud computing, and the MapReduce programming paradigm, to distribute subtasks containing part of the
    mutation analysis task across Amazon EC2~\cite{amazon_web_services_amazon_2015} computing nodes.  We empirically analyse two different work distribution strategies for our tool. We compare the two strategies' scalability 
    and overall performance on three open source software projects.  Furthermore, we compare both strategies with a state-of-the-art non-distributed tool, PiT~\cite{coles_pit_2015}.  We analyze
    the runtime behaviour of the system to identify opportunities for more accurate division of
    mutation analysis tasks.  In our discussion, based on our results, we present concepts for a more efficient work 
    distribution model, and identify another opportunity for optimization of mutation analysis within a continuous integration workflow---incrementally updating mutation scores as
    the software and test suite evolves.

\section{Preliminaries} 

\subsection{Mutation Analysis Preliminaries}
Acree et al.~\cite{acree_mutation_1979} define a process for using mutation analysis to assess a test suite's ability to detect actual faults in a program:
\begin{enumerate}
\item A set of \emph{mutant} programs, $M$, is generated by applying \emph{mutation operators} to a program $P$. Each mutant $m \in M$ differs from $P$ by the inclusion of some small \emph{fault} such as the replacement of a $>$ operator with a $\geq$ operator, or a logical \texttt{AND} with a logical \texttt{OR}. This is intended to change the behaviour of $P$ while maintaining a syntactically correct program $m$. The mutation (\ie{the injected fault}) is defined by the \emph{mutation operator} which, when applied to $P$, produces a subset of $M$. Examples of mutation operators in PiT, a modern mutation testing system for Java, are shown in Table~\ref{tab:mutation-operators}.
\begin{table}
  \centering
  \scalebox{0.7}{%
  \begin{tabular}{lcc}
    \hline
    & \multicolumn{2}{c}{\textbf{Program Statement}} \\
    \cline{2-3}
    \rule{0pt}{2.5ex}
    \textbf{Operator} & \textbf{Before Mutation} & \textbf{After Mutation} \\
    \hline
    \texttt{INVERT NEGS} & \texttt{-i} & \texttt{i} \\
    \texttt{MATH} & \texttt{a + b} & \texttt{a - b} \\
    \texttt{RETURN VALS} & \texttt{return true} & \texttt{return false} \\
    \hline
  \end{tabular}
  }
  \caption{Examples of Mutation Operators in the PiT Mutation Testing System~\cite{coles_pit_2015}.}
  \label{tab:mutation-operators}
\end{table}

\item The original program $P$ is executed against a set of test cases $T$. If $P$ does not pass all test cases $t \in T$, then the faults injected during mutation analysis cannot be differentiated from legitimate faults in the program and hence, the analysis cannot continue.

\item Each mutant $m \in M$ is executed against each test case $t \in T$. If any test case $t$ \emph{fails} when executed against $m$, the mutant $m$ is said to have been \emph{killed}; otherwise  $m$ is said to be \emph{alive} and a gap in $T$ has been identified. It should be noted that a mutation will occasionally produce no observable change in the output of a program on any input. These \emph{equivalent mutants} are a well-documented problem in mutation testing \cite{budd_two_1982,jia_analysis_2011}, and are not the subject of the present work.

\item If the number of killed mutants is $k$, and the number of live (non-equivalent) mutatnts is $l$, the \emph{mutation adequacy score} (alternatively, mutation score), $S_T$, of the test suite $T$ is then defined as:
\[
S_T = \frac{k}{l},
\]
Some mutation analysis systems use an alternative definition for the mutation score:$\frac{k}{k+l}$, the ratio of killed mutants to the total number of testable mutants created. In either case, a higher mutation adequacy score indicates a greater ability of a test suite to detect faults in a program. 
\end{enumerate}

\subsection{MapReduce Programming Model}
\label{sec:map-reduce}
MapReduce is a programming model that is well-suited to processing large datasets on distributed systems \cite{dean_mapreduce:_2008} (that is, parallelized computation where the compute nodes do not share access to the same physical main memory~\cite{ghosh_distributed_2014}). MapReduce implementations provide two functions---\emph{map} and \emph{reduce}. The \emph{map} function takes as input a collection of subsets of some large dataset to process and computes a collection of partial results. The \emph{reduce} function aggregates each partial result into a final result. MapReduce programs are naturally parallelisable---each invocation of the \emph{map} function can be executed on any available processing node. The \emph{reduce} function then provides a simple way to compute a final result from a collection of partial results without requiring coordination between each \emph{map} task.  MapReduce allows the complexities of dynamic task scheduling and node failure to be handled transparently by the implementation \cite{dean_mapreduce:_2008}.  
\emph{Hadoop}~\cite{white_hadoop:_2012} is an open-source framework that provides reliable distributed computing constructs, including MapReduce functionality \cite{dean_mapreduce:_2008}.  Hadoop allows for efficient processing of extremely large datasets on distributed systems formed from commodity hardware. Spark~\cite{zaharia_spark:_2010} is a newer framework for cluster computing, that also supports MapReduce but adds facilities for distributed shared memory.

When designing MapReduce programs, it is important to consider the granularity of each map task--- that is, how much of the original input data is partitioned to each subtask. If the size of each partition is too small and the number of subtasks too large, the system is likely to incur overhead in scheduling and network communication. Conversely, if the partitions are large and the number of subtasks is small, then imbalances in subtask processing time will result in delays while a minority of nodes complete large subtasks.

\section{A Model for Distributed Mutation Testing}
\label{sec:distributed-model}

\subsection{Introduction}
In this section, we present a simplified model for mutation testing using the MapReduce abstraction. We describe a basic system architecture for this purpose, define key terms used in the remainder of this paper, and attempt to justify the assumptions we've made in simplifying this problem.

\subsection{System Architecture}
We assume a simple distributed system where a singular \emph{master node} coordinates work across some cluster of \emph{worker nodes}. Symmetric communication channels exist between the master node and each worker node, but no communication is required between worker nodes. Communication between worker nodes is not required for mutation analysis, and its absence simplifies the computation required to coordinate large parallel jobs. An annotated diagram of this architecture is presented in Figure~\ref{fig:work-distribution:definition-diagram}, and key terms are defined below:
\begin{figure}
  \centering
  \includegraphics[width=\linewidth]{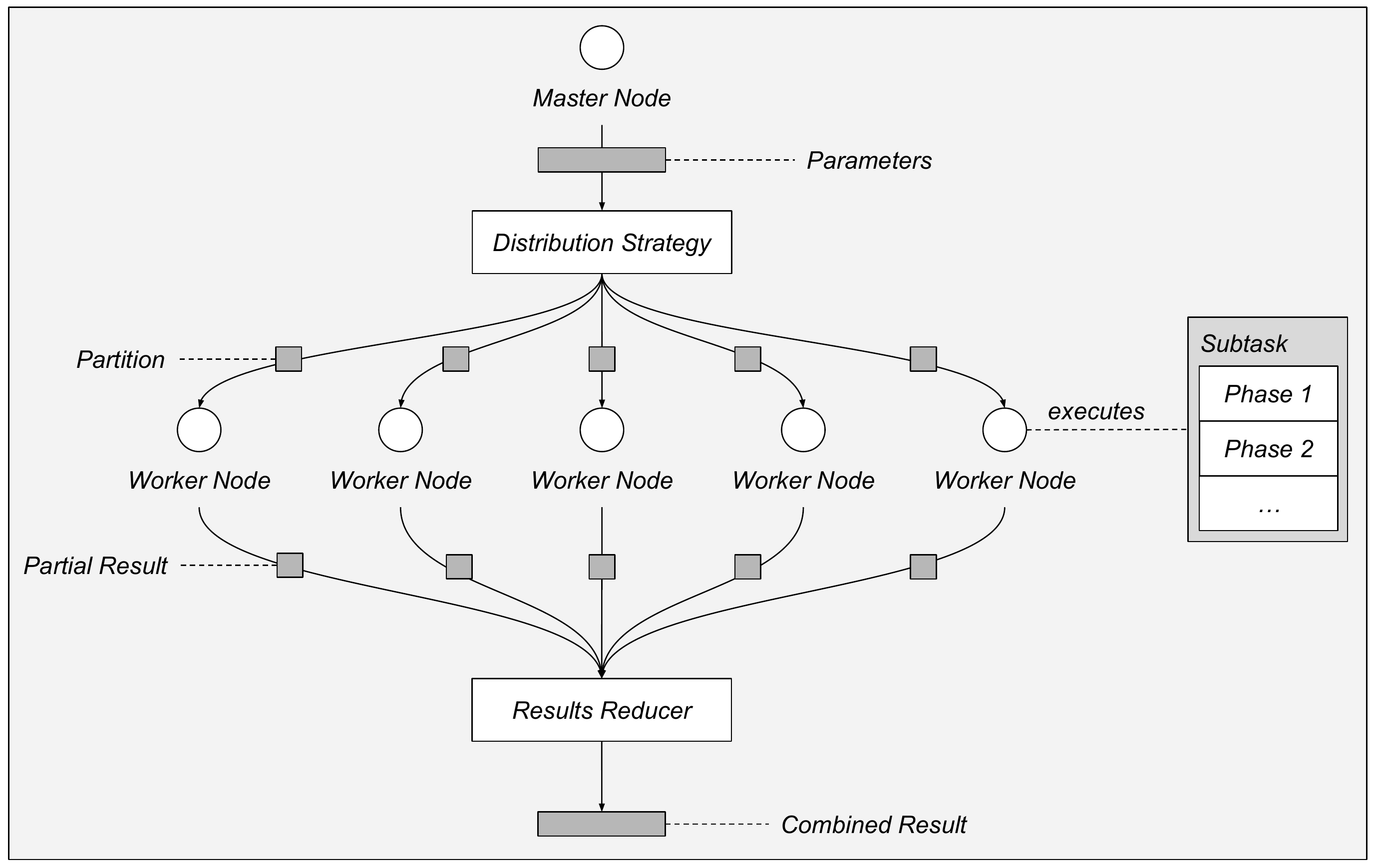}
  \caption[Distributed Mutation Testing Definitions]{Key definitions in our distributed mutation testing model.}
  \label{fig:work-distribution:definition-diagram}
\end{figure}
\begin{description}
\item[Master Node]
One node in the cluster that coordinates and partitions work amongst the available worker nodes. This is the machine that a user will interact with to deploy jobs to the cluster.
\item[Worker Node]
One of many processing nodes in the cluster that execute tasks as assigned by the master node.
\item[Parameters]
The inputs to a mutation testing task---a set of classes, tests, and mutation operators.
\item[Partition]
A subset of the parameters supplied to a worker node, that is, a \emph{partial input} from which to compute a partial result.
\item[Distribution Strategy]
The method by which parameters are partitioned amongst available worker nodes. Some components may be mapped to worker nodes one-by-one, and others may be broadcast to all worker nodes. For example, one possible distribution strategy would broadcast all tests and mutation operators amongst all available worker nodes, then map individual classes to each partition.
\item[Subtask]
A partial mutation analysis task that is executed on a worker node. A subtask takes a partition as input, and computes a partial result.
\item[Phase]
A discrete stage or component inside a mutation analysis task or subtask (\eg{\emph{generate mutants}, or \emph{execute mutants}}).
\item[Partial Result]
The output of a partial mutation analysis task executed by a worker node. This will contain the status of the mutants generated by the mutation analysis subtask. \item[Combined Result] The aggregation of partial results from all worker nodes. This is the final result of the mutation analysis.
\end{description}

\subsection{Parallelisable Components}
\label{sec:work-distribution:parallelisable-components}
There are virtually no data dependencies or synchronisation constraints within each stage. Mutants can be generated from different mutation operators and different classes concurrently, and once generated, executed and evaluated independently. This immediately offers two opportunities for parallelization:
\begin{enumerate}
\item Subsets of the set of all mutants $M$ can be generated concurrently by partitioning available classes and mutation operators amongst available processing nodes; 
\item The set of all mutants $M$ can be partitioned amongst available processing nodes and executed concurrently.
\end{enumerate}
Hybrid approaches are also possible, in which a subset of $M$ is both generated and executed on a processing node from some partition of classes and mutation operators.

\subsection{Parameters}
A mutation analysis task operates on a program $P$, where $P$ is a collection of classes and tests, and is defined by the following parameters:
\begin{description}
\newcommand{\litem}[2]{\item[#1 \enspace $#2$ \enspace]}
\litem{Classes}{C}
The subset of classes in $P$ from which mutants should be generated. Typically all classes in $P$ will be included, but specific classes may be excluded in some cases (\eg{to reduce long running times by analysing only a specific subsystem in $P$}).
\litem{Tests}{T}
The subset of test cases in $P$ to use during the execution of mutants, \ie{the test cases to run against the mutants generated from $P$}. As is the case with $C$, we may wish to manually exclude irrelevant test cases to prevent long running times.
\litem{Mutation Operators}{O}
The set of mutation operators to be applied to $P$ when generating mutants. In techniques such as \emph{Selective Mutation}~\cite{offutt_experimental_1993}, only the most useful mutation operators are applied to a program. We may also wish to restrict these for a specific domain (\eg{mutation operators targeting arithmetic operators are perhaps more applicable in arithmetic-heavy programs than in others}).
\end{description}
The output of a mutation analysis task is a set of results $R$.  They contain the mutation adequacy score for the program $P$, and may contain other information such as live mutants and timing data.

\subsection{Phases}
\label{sec:model:phases}
\begin{figure}
\centering
\includegraphics[width=0.5\linewidth]{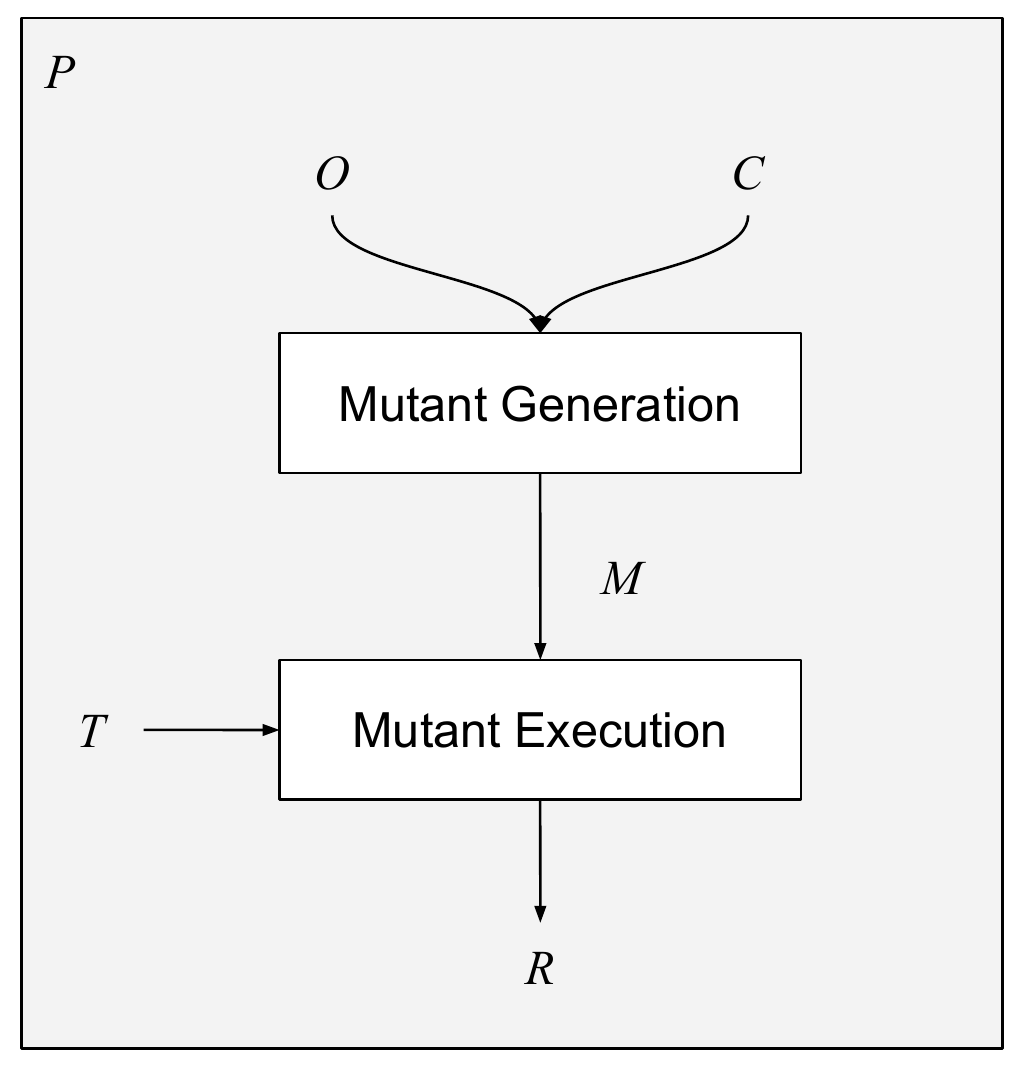}
\caption[Mutation Testing Process]{An idealized model of the mutation testing process.}
\label{fig:work-distribution:simplified-mutation-testing-process}
\end{figure}

There are two main \emph{phases} to complete during the mutation analysis of a program $P$, as shown with inputs and outputs in Figure~\ref{fig:work-distribution:simplified-mutation-testing-process} :
\begin{description}
\item[Mutant Generation]
The set of mutation operators $O$ are applied to the set of classes $C$ to produce a set of mutants $M$; followed by
\item[Mutant Execution]
The set of test cases $T$ are executed against each mutant $m \in M$. The outputs of this phase (\ie{which mutants were detected by a failing test case and which remain alive}) comprise the set of results $R$.
\end{description}

\subsection{Assumptions}

\label{sec:work-distribution:shared-program-data}
Both the generation and execution phases operate on the entire program $P$. Although it is theoretically possible to partition a program by class and make only the components in the execution path of tests covering that partition available to each processing node, we assume that all of $P$ is available to each individual processing node at all times. All classes $c \in C$ and tests $t \in T$ then simply denote an \emph{identifier} for some component of $P$ (\ie{a class name})---not the raw data comprising that class or test.   Therefore, the entire program $P$ must be transferred to each worker node.

The overhead incurred is unlikely to be practically important.  An example of a very large, monolithic
software package is the Chrome web browser.  On the Ubuntu operating system, the binary packages
take up around 50MB, which will take around one second to transfer across a  local network
of the capacity typically used by cloud computing providers. The largest program used in our experiments, described in \S\ref{sec:empirical-evaluation}, is \emph{JodaTime}---a Java program with almost 300 combined classes and test classes. The jar file used to archive and transfer this program between processing nodes is smaller than 1.5 MB.  Compared to the tens of minutes required for mutation analysis of \emph{JodaTime} even when using a 16-core cluster, even
a one second delay is insignificant.

\section{A Prototype Distributed Mutation Testing System }
\subsection{Introduction}
This section describes an implementation of the abstract distributed mutation testing system described in \S\ref{sec:distributed-model}. This system is subsequently used to evaluate the suitability of distributed computing in mutation analysis in \S\ref{sec:empirical-evaluation}.

\subsection{Tools and Frameworks} 
Our system is built largely on existing tools and frameworks. While this imposed a number of limitations, it also allowed us to build and evaluate a working prototype quickly, to inform later, more sophisticated implementations. 

Java was chosen as the implementation and target language for the prototype.  This was due to the industrial relevance of Java and the availability of actively-developed mutation analysis and distributed computing tools which could be utilized for prototype construction.  

After some preliminary investigation of Hadoop~\cite{white_hadoop:_2012}, Spark~\cite{zaharia_spark:_2010} was chosen as the distributed computing platform for our prototype. This was mainly due the ease at which local clusters can be setup in Spark for testing and debugging, and the support for creating clusters on Amazon's~EC2 service; Spark's fast distributed shared memory was not actually required for our application. Note that
Spark treats multiple processor cores on the same system as, in effect, independent nodes.  Therefore, in our experiments, where we used twin-core Amazon compute nodes, we describe the size of our clusters in ``cores''.

PiT~\cite{coles_pit_2015} was chosen as the mutation engine for our prototype. It is actively developed, open-source, written in Java, already runs in parallel on a single machine, and claims to represent the ``state-of-the-art'' in mutation testing~\cite{coles_pit_2015}---a claim we consider reasonable after brief experimentation with alternatives. Its existing support for local parallel mutation testing provided us with some confidence in its ability to support distributed parallel processing, and its self-proclaimed status as ``state-of-the-art'' provides a useful performance baseline for evaluating our results.  In retrospect, some aspects of PiT proved less than ideal for our purposes, as we will discuss in \S\ref{sec:limitations}.

\subsection{Architecture}
We attempted to realise the model described in \S\ref{sec:distributed-model} as faithfully as possible. The prototype is written in Java and is invoked as a command-line utility that allows users to specify:
\begin{description}
\item[A Java Program] comprised of a set of classes and tests to use as input to the mutation analysis;
\item[A Set of Mutation Operators] to apply during mutant generation. Since our system uses PiT as the mutation testing component, we have the same set of available mutation operators as PiT (seven are used by default~\cite{coles_pit_2015-1}); and
\item[A Distribution Strategy] that defines how the required work is partitioned to worker nodes in the cluster. Available distribution strategies in our prototype are visualised in Figure~\ref{fig:distribution-strategies}.
\begin{figure}
  \centering
  \begin{subfigure}{\linewidth}
    \includegraphics[width=\linewidth]{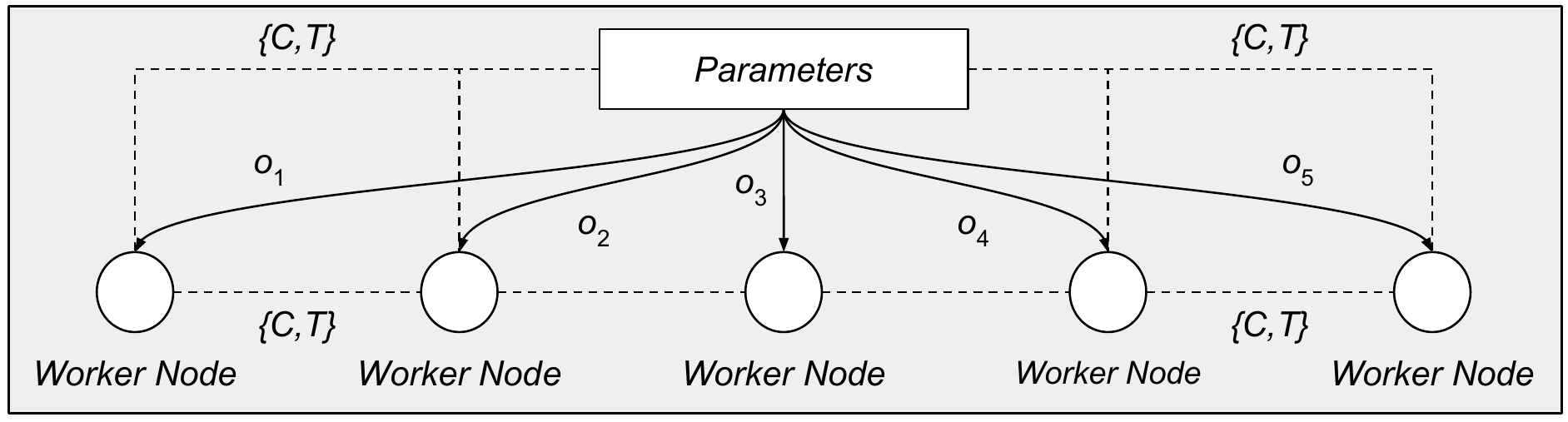}
    \caption{\emph{Parallelise by Mutation Operator}---All Classes and Tests are broadcast to every worker node, and mutation operators are mapped out individually.}
  \end{subfigure}
  \par\medskip
  \begin{subfigure}{\linewidth}
    \includegraphics[width=\linewidth]{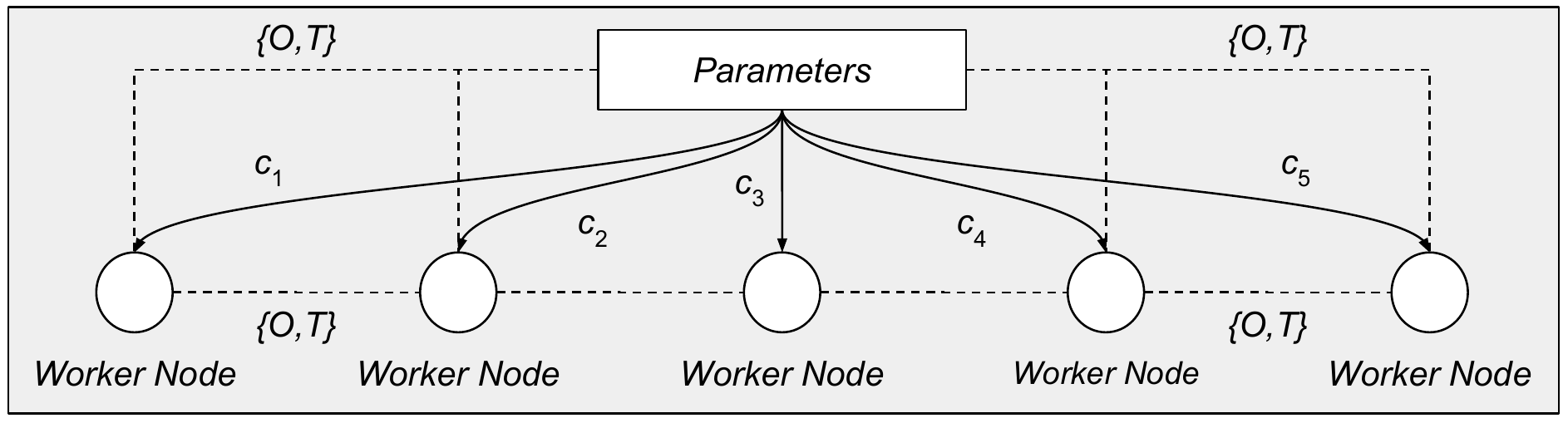}
    \caption{\emph{Parallelise by Class}---All mutation operators and tests are broadcast to every worker node, and classes are mapped out individually.}
  \end{subfigure}
  \caption{Distribution Strategies in our prototype.}
  \label{fig:distribution-strategies}
\end{figure}
\end{description}

The mutation testing process is then distributed across a cluster hosted on Amazon's EC2 service~\cite{amazon_web_services_amazon_2015}.  In our experiments, we used \texttt{m1.large} nodes with two processor cores and 7.5GB
of RAM (allocated equally to the two cores by Spark).  The single processing node experiments were
        performed by deploying the prototype to a single node in serial local
        debug mode. All other cluster sizes were achieved by initialising Spark with varying numbers
        of worker nodes.  Note that while Spark avoids resending shared data to each individual core
        on the node, for scheduling purposes Spark treats each processor core allocated to it as an independent processing node.  Because of these factors, 
        in \S\ref{sec:empirical-evaluation} we describe cluster sizes by the number of processor cores.
        
The output of the program is a JSON dictionary containing a collection of partial results for each subtask, as well as an aggregated result representing the combination of all partial results. These partial results contained timing data and other diagnostic information which are not necessary for an industrial tool, but allowed us to conduct the experiments described in \S\ref{sec:empirical-evaluation}.

\begin{figure}
\centering
\includegraphics[width=\linewidth]{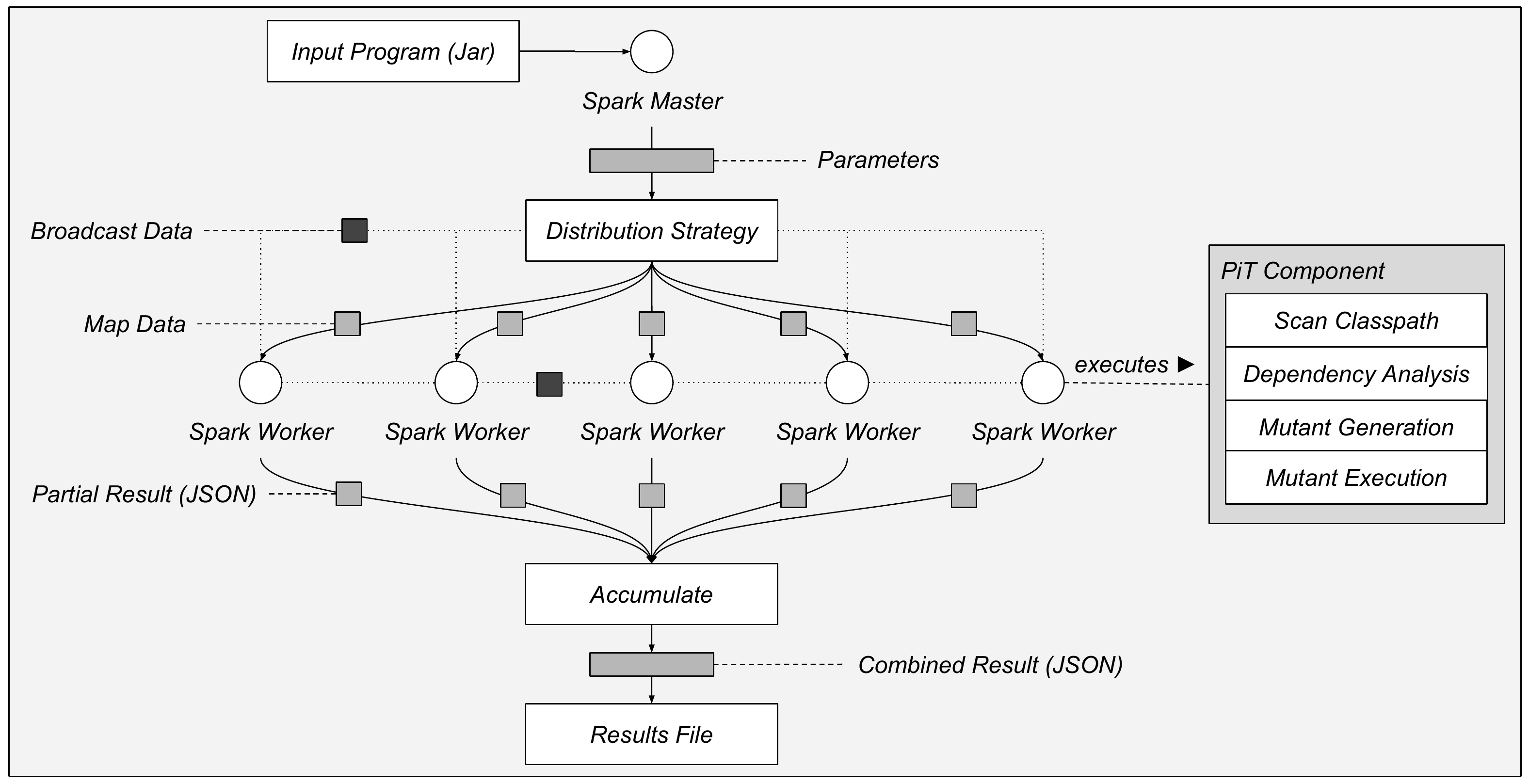}
\caption[Prototype Architecture]{High-level architectural overview of our distributed mutation testing prototype.}
\label{fig:prototype-implementation:architecture}
\end{figure}

We have modified  PiT  to function as a programmable component that can be initialised with mutation testing parameters (\ie{classes, tests, and mutation operators}) dynamically, and executed to produce a JSON object containing the results of the analysis, along with other metrics such as durations of various phases and subject class sizes. We execute subtasks on Spark workers as map tasks. Essentially, we:
\begin{enumerate}
\item Provide the prototype with the input program as a jar file (containing both classes and tests);
\item Use Spark's broadcasting facilities to share the input program jar file with the worker nodes in the cluster;\footnote{Note that Spark employs efficient broadcast algorithms and protocols specifically for this purpose~\cite{apache_foundation_spark_2015}.}
\item Use a combination of Spark broadcast and map tasks to send some partition of classes, tests, and mutation operators to each processing node (defined by the choice of distribution strategy);
\item Execute a mutation analysis subtask on each Spark worker node to produce a partial result; 
\item Combine each partial result at the Spark master via a simple reduce function; then
\item Write the partial and combined results to an output file for further analysis.
\end{enumerate}
Our implementation architecture is summarised in Figure~\ref{fig:prototype-implementation:architecture}.

\subsection{Additional Phases}
\label{sec:prototype-implementation:additional-phases}
PiT's actual mutation testing process is slightly more complicated than the process we described in Figure~\ref{fig:work-distribution:simplified-mutation-testing-process}.  It introduces two additional phases not depicted in our abstract model:
\begin{description}
\item[Scan Classpath]
The supplied class and test names are located on the classpath, that is, the actual data
making up these parameters is resolved from the identifiers provided by the user; and
\item[Dependency Analysis]
\emph{Dependency Distance} between classes under analysis and other classes in the
program is computed. This is defined as follows---if Class $A$ calls a method from Class
$B$, then Classes $A$ and $B$ have a dependency distance of 1. If Class $B$ calls a method
from Class $C$, the classes $A$ and $C$ have a dependency distance of 2.
\end{description}
PiT provides an option to ignore tests beyond a given dependency distance from a target class.
This is intended as a strategy for reducing runtime by ignoring potentially irrelevant tests.
We have configured PiT to ignore this feature for our experiments, but the dependency analysis
is still executed regardless. This introduces some overhead in our subtasks,  but 
 the modifications required to remove the overhead could not 
have been completed in the time available for the project.

\subsection{Limitations}
\label{sec:limitations}
\subsubsection{Mutant Generation}
\label{sec:prototype-implementation:mutant-generation-limitation}
We use PiT effectively as a \emph{black box}. We provide it with input
parameters for a mutation analysis task (or subtask), and we process the output (\ie{a
collection of mutants that were killed, and those that are still alive}). This makes it
difficult to exploit parallelism that may exist \emph{inside} PiT's mutation testing process.
For example, we are unable to generate mutants serially at the master node and dispatch them
to processing nodes individually---all mutants are generated and executed within the same
subtask. Correcting this may have allowed for more advanced distribution strategies, but would
have required significant programming effort. 

\subsubsection{Constant Overheads}
\label{sec:prototype-implementation:constant-overheads} 
PiT implements some optimisations intended to improve runtime performance in a non-distributed
environment. This includes a \emph{coverage analysis} phase in which the test suite is
executed without mutation to determine unreachable sections of code, and a \emph{dependency
analysis} phase, discussed in
\S\ref{sec:prototype-implementation:additional-phases}. As discussed, we did 
not have the time to modify PiT to disable these steps,
and so they are repeated for every subtask at every processing node. We take this into consideration when evaluating our system.

\subsubsection{Economic Constraints} 
\label{sec:prototype-implementation:economic-constraints}
Our cluster was hosted on Amazon's EC2 service. This is
especially appropriate considering that it is our goal to eventually provide mutation testing
as a service in the cloud. However, we were unable to run our experiments on Amazon's free tier, 
and all usage costs were paid by the authors.  This restricted the size of the cluster 
we were able to use for our experiments, and restricted the number of repetitions of experiments
that could be completed to confirm that measurements did not vary significantly between trials.

\section{Empirical evalution} 
\label{sec:empirical-evaluation}

We conducted empirical evaluations of our tool to answer two main questions:

\begin{enumerate}
 \item Can a distributed approach perform mutation analysis faster than a non-distributed approach---and, if so, how efficiently can it be scaled?
 \item Can we accurately predict the amount of work required to perform some fraction of the mutation analysis task so that work can be divided up between computational nodes to maximise utilization and thus efficiency?
\end{enumerate}

Research question 1 directly addresses the viability of our general approach to distributed mutation analysis.  Research question 2 addresses how an optimized tool might 
be constructed to achieve the best performance and lowest resource utilization.   We conducted three experiments, measuring different aspects of the performance of the prototype system:

\begin{description}
\item[Experiment 1] measured the execution times of the prototype system using different work distribution strategies and cluster sizes.  This was primarily intended to address research question 1.

\item[Experiment 2] investigated the relationship between the size of individual classes within the software under test, the number of mutants created within those classes, and the time taken to process the created mutants.  This experiment addresses research question 2.

\item[Experiment 3] examined the number of mutants generated by applying different mutation operators to the software under test, to determine whether the proportions were consistent across projects.  This experiment addresses research question 2.
\end{description}

\subsection{Experimental subjects}
      \label{subsec:experimental-subjects}
      To conduct our three experiments, we required experimental subjects---representative software systems for \emph{input} to our experiments (\ie{to mutate and analyse}).
      We required experimental subjects that were implemented in Java (as PiT only supports Java) and used the JUnit unit testing framework~\cite{gamma_junit_2006}.  We also preferred open source projects to enable easier replication
      of our results by future researchers.
   
      For our study, we chose three open source libraries as input to our prototype, whose source repositories are available from the online project repository site \emph{GitHub}~\cite{dabbish_social_2012}:
      \begin{description}
        \newcommand{\pitem}[2]{\item[#1]#2 \enspace}
        \pitem{Gson}{\cite{singh_gson_2008}}
          A popular JSON parsing library;
        \pitem{Joda-Time}{\cite{colebourne_joda_2010}}
          A de facto standard replacement library for Java's Date and Time utilities; and
        \pitem{Commons Math}{\cite{apache_foundation_apache_2014}}
          A collection of mathematical utilities. Commons Math is a very large project, and serial
          mutation testing benchmarks for the entire project take hours to complete on our machines.
          In the interest of keeping our experiment runtimes reasonable, only the \texttt{geometry} package of the Commons Math project (and its
          associated tests) were used.
      \end{description}
      The size of these projects are shown in Table~\ref{tab:results:project-sizes}.
      \begin{table}
        \centering
        \scalebox{0.7}{%
        \begin{tabular}{lccccc}
          \hline
                           &                  &                  & \textbf{Mutatable}      & \textbf{Executable} & \\
                           &                  & \textbf{Test}    & \textbf{Lines of}      &  \textbf{Test}   & \textbf{Mutants}                 \\
          \textbf{Project} & \textbf{Classes} & \textbf{Classes} & \textbf{Code}  & \textbf{Cases} & \textbf{Generated} \\ \hline
          gson & 44 & 78 & 2567 & 943 & 1585 \\
          joda-time & 161 & 133 & 11378 & 12480 & 8037 \\
          commons-math & 71 & 47 & 4679 & 547 & 3745 \\
          \hline
        \end{tabular}
        }
        \caption[Project Sizes]{The sizes of each project used as input to our experiments.}
        \label{tab:results:project-sizes}
      \end{table}
  

\subsection{Method}
\label{subsec-method}
\subsubsection{Experiment 1}
      Using our prototype, we perform mutation analysis of:
      \begin{itemize}
        \item Each program listed in \S\ref{subsec:experimental-subjects}; using 
        \item Each distribution strategy described in Figure~\ref{fig:distribution-strategies}; with 
        \item 1, 2, 4, 8, 12, and 16 processor cores available.  These configurations are summarised in 
        Table~\ref{tab:prototype-analysis:spark-configuration}.
      \end{itemize}
      \begin{table}
        \centering
        \scalebox{0.7}{%
        \begin{tabular}{cl}
          \hline
          \textbf{Nodes} & \textbf{Spark/EC2 Configuration} \\
          \hline
          1 & Serial Spark deployed to master node \\
          2 & Distributed Spark with one worker node \\
          4 & Distributed Spark with two worker nodes\\
          8 & Distributed Spark with four worker nodes \\
          12 & Distributed Spark with six worker nodes \\
          16 & Distributed Spark with eight worker nodes \\
          \hline
        \end{tabular}
        }
        \caption[Experiment Cluster Configurations]{Spark and EC2 configurations used for each
        cluster size used in our experiments.}
        \label{tab:prototype-analysis:spark-configuration}
      \end{table}
      
      To compare the performance with a non-distributed (but locally parallelised) implementation, we also used an unmodified version of PiT, configured to take advantage of both available cores, to perform mutation analysis on a single node.  
      
      In all experiments, we apply the set of \emph{default} mutation operators as defined by PiT~\cite{coles_pit_2015-1}.
      
      Therefore, the \emph{independent} variables for Experiment 1 are:
      \begin{itemize}
       \item The software under test;
       \item The distribution strategy;
       \item Whether our modified distributed tool or standard PiT was used to perform mutation analysis; and
       \item The number of processor cores allocated.
      \end{itemize}

      The \emph{dependent} variable was the clock time (measured in milliseconds using the system clock on the master node) required to complete mutation analysis for the software under test.
      
      For the 16-node distributed case, and the standard PiT case, we repeated measurements three times.  The variation between experimental runs was insignificant.  
      \ifx\appendixversion\undefined
      Full details are presented
      in an extended technical report version of this paper~\footnote{identifying citation removed for review}.
      \else
      See the appendix for full details.
      \fi
      
      $D_P$ is the duration of the analysis when deployed to $P$ processing nodes, and $D_1$ is
    the duration of the analysis when executed on one processing node.
    
\subsubsection{Experiment 2}

We performed mutation analysis using the prototype tool, using individual classes as partitions.  We recorded 
the number of mutatable lines of code in each partition.  The statistic is provided by PiT, and excludes method headers, whitespace, and lines consisting entirely of non-mutatable constructs such as braces.  We also recorded the execution time, provided by PiT, for mutants relating to that class, subdivided into four phases:

\begin{description}
 \item[Classpath Scan] Determine the location of all Java classes in the artifact and test suite necessary to run the tests;
 \item[Dependency Analysis] Determine the shortest invocation path between all pairs of classes in the system.  As noted in \S\ref{sec:prototype-implementation:constant-overheads}, this was unused overhead, but was not feasible to disable in the time available;
 \item[Mutant Generation] Generate mutants by applying mutation operators to the input program; and
 \item[Mutant Execution] Execute each mutant against the test suite.
\end{description}

\subsubsection{Experiment 3}

We performed mutation analysis using the prototype tool, partitioning by mutation operator.  We recorded the number of mutants created in the software under test using each default mutation operator
available in PiT, as well as the time required to complete the partial mutation analysis for the mutation operator.





\section{Results}

\subsection{Experiment 1}

\begin{figure*}
      \centering
      \includegraphics[width=\linewidth]{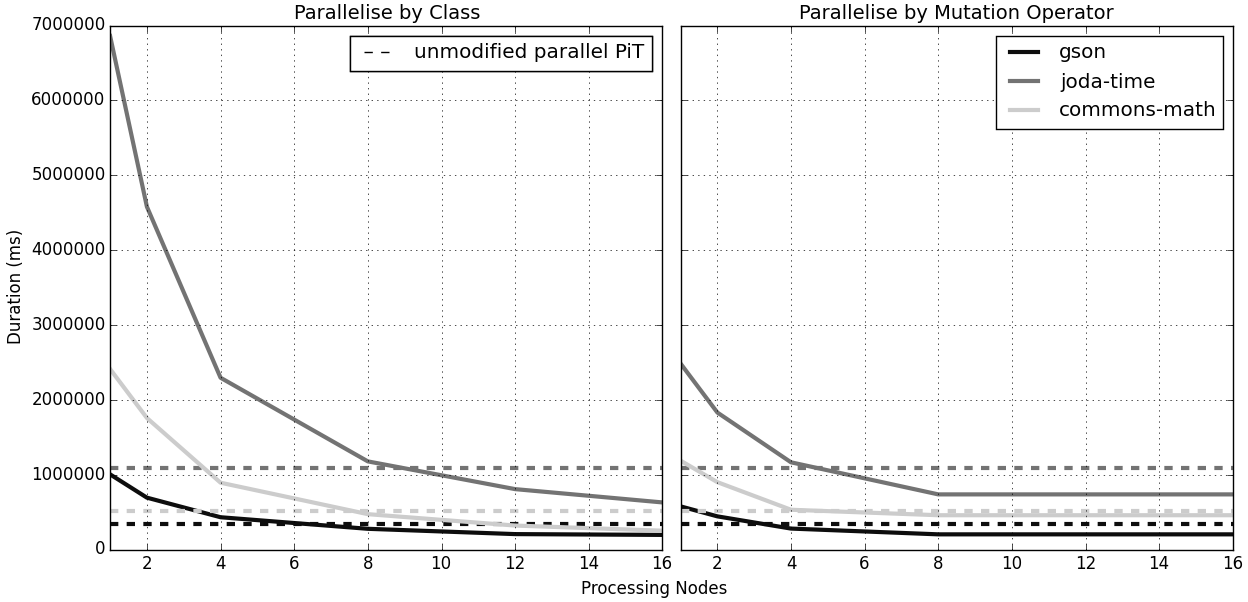}
      \caption[Scalability of Distribution Strategies (Durations)]{Durations of distribution
      strategies as cluster size increases.}
      \label{fig:results:scalability-durations}
      \end{figure*}

Figure~\ref{fig:results:scalability-durations} shows the clock time taken to complete mutation analysis, on each of the three experimental artifacts, 
 using the two distribution strategies for cluster sizes of between 1 and 16 cores. The time taken for an unmodified version of PiT running two threads on a single twin-core node to complete mutation analysis 
 for each of the experimental artifacts is indicated by the horizontal line of the corresponding shade.
 
The figure shows that the Parallelise by Mutation Operator strategy (in which each node performed all mutant generation and execution for one mutation operator) performed better on smaller clusters, but as there are only seven default mutation operators, adding additional nodes to the cluster after this did not reduce the time taken.  By contrast, the Parallelise by Class strategy (in which partitions consisted of all mutants for a single class) was far slower on small clusters, but execution time continued to reduce as the cluster was expanded up to sixteen
cores. With 16 cores available, the clock time to execute the Parallelise by Class strategy was lower than for Parallelise by Mutation Operator.

With a sufficiently large cluster, both strategies were able to complete mutation analysis faster than non-distributed PiT, though the speedup was far from the theoretical maximum. With 16 cores, the Parallelise by Class strategy was able to complete
mutation analysis 40\% faster than non-distributed PiT for Gson, 42\% faster for  Joda-Time, and 50\% faster for Commons Math. 
 
\subsection{Experiment 2}
\begin{figure}
      \centering
      \includegraphics[width=\linewidth]{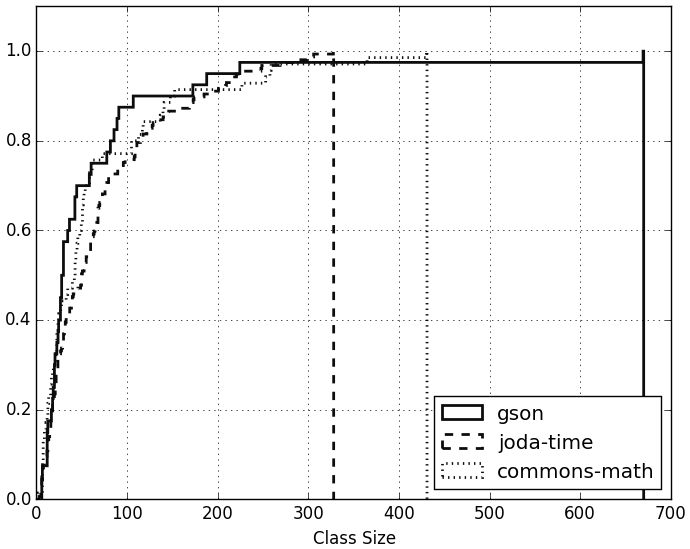}
      \caption[Class Size Distribution]{Distribution of class sizes for each project.}
      \label{fig:results:class-size-distribution}
    \end{figure}

Figure~\ref{fig:results:class-size-distribution} shows the distribution of class sizes for each project.  We found less variance than expected. Of
    the 276 classes analysed, the majority are between approximately 75 and 150 mutatable lines of code.  
    
\begin{figure}
      \centering
      \includegraphics[width=\linewidth]{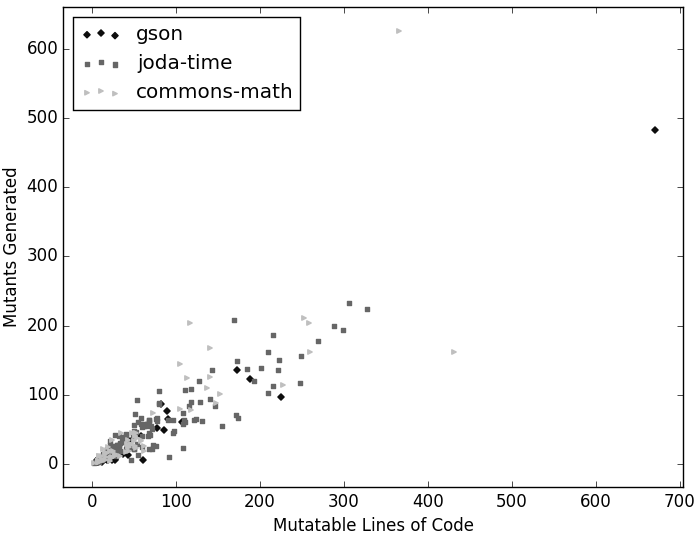}
      \caption[Mutants Generated by Class Size]{Mutants generated by class size.}
      \label{fig:results:mutants-generated-by-class-size}
    \end{figure}

The relationship between the size of the class and the number of mutants generated is shown in Figure~\ref{fig:results:mutants-generated-by-class-size}.  Across all projects, there is a strong linear relationship between class
size and mutants generated.

\begin{figure}
      \centering
      \includegraphics[width=\linewidth]{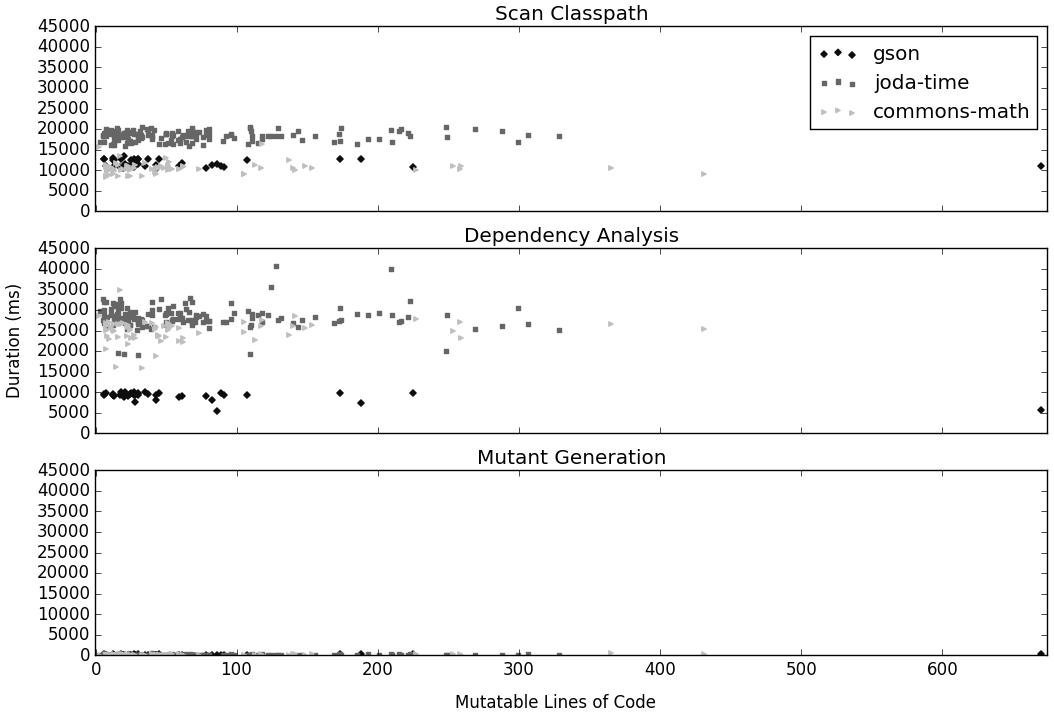}
      \caption[Durations by Mutatable Lines of Code]{Durations by mutatable lines of code.}
      \label{fig:results:durations-by-mutable-lines-of-code}
    \end{figure}

Figure~\ref{fig:results:durations-by-mutable-lines-of-code} shows the duration of time spent in each phase of the mutation analysis process using our prototype tool, when using the Parallelise by Class strategy. Classpath scanning, dependency analysis, and mutant generation were essentially unaffected by the size of the class being processed.  This is unsurprising for classpath scanning and dependency analysis, as they depend on the properties of
the entire project, not just the specific class.  Mutant generation, by contrast, is fast enough compared to the other phases that its computational cost can be ignored.  

\begin{figure}
      \centering
      \includegraphics[width=\linewidth]{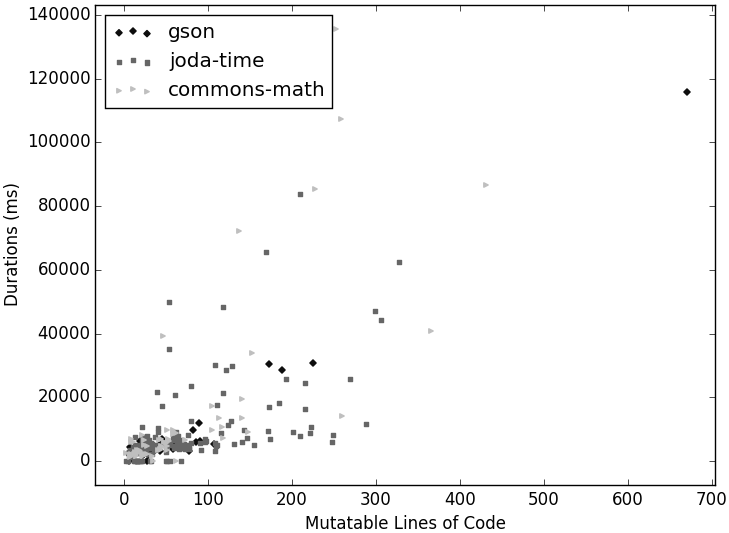}
      \caption[Mutant Execution Time by Class Size]{Mutant execution time by class size.}
      \label{fig:results:mutant-execution-time-by-class-size}
    \end{figure}

There is a moderately strong linear relationship ($r^2 = 0.498$) between class size and the time taken to complete mutant execution for all mutants relating to that class, as shown in Figure~\ref{fig:results:mutant-execution-time-by-class-size}.

\subsection{Experiment 3} 

\begin{figure}
      \centering
      \includegraphics[width=\linewidth]{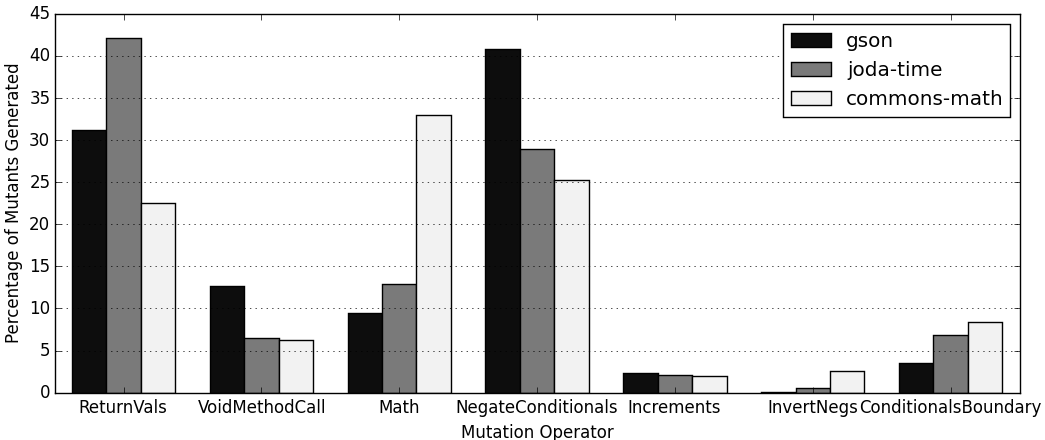}
      \caption[Mutants Generated by Operator]{Mutants generated by operator.}
      \label{fig:results:mutants-generated-by-operator}
    \end{figure}

Figure~\ref{fig:results:mutants-generated-by-operator} shows the number of mutants generated by applying each of the mutation operator classes to our experimental subject programs.  There is no clear pattern to the proportion
of mutants across projects.

\subsection{Threats to validity}


We have reasonable confidence in the internal validity of the results.

It is possible, but unlikely, that our prototype tool was not performing mutation analysis as described in the paper due to bugs. Comparisons between our modified tool and unmodified PiT suggest that the tool produced equal quantities of mutants when applied to the same Java source files.

Faults in the (new) code used to distribute test cases across nodes is a more likely source of invalid data.  To mitigate this threat, an extensive unit test suite was devised for all the new code in the prototype tool, which
caught a number of bugs during development.

It is possible that timing data collected from within PiT or by the Apache Spark cluster software may have had inaccuracies.  These are unlikely to be substantial given the relatively long durations measured in our experiments.  Variations in network load may also have caused timing variations.  The lack of variation we observed in our (limited) repeated observations suggest that this is likely to be relatively minor.




There are a number of limits to the external validity of our results.  

The tool which we used for our experiments was a proof-of-concept prototype, and its performance is unlikely to be representative of a properly engineered production system.  However, given the nature of the limitations of the prototype (notably, the amount of unnecessary work performed and the inefficient partitioning schemes) that any more advanced system is likely to be faster in absolute terms, and exhibit better scalability rather than worse.

Our tool was restricted to software written in Java and tested using the JUnit testing framework.  The overheads from the different parts of the mutation analysis process are highly dependent on the details of how mutants are inserted, and the overheads of the testing framework---as such, optimal parallelization strategies may vary for programs implemented with a different toolchain.

Finally, time limitations meant that we were only able to evaluate our tool using three experimental subjects, all of which were open source utility libraries of one kind or another.  With regards to Experiment 3, where
significant differences in the proportion of mutants created by applying the various mutation operators to the three experimental subjects,  we lack enough data to say whether these differences represented rare outliers or wide variation.

%
%
%
%
%

\section{Discussion}

Our results for Experiment 1 demonstrate that the answer to research question 1 is in the affirmative: a distributed, cloud based approach to mutation analysis can outperform a non-distributed system.  Our proof-of-concept prototype, despite known inefficiencies, was able to significantly outperform the non-distributed tool it was based on.  Parallelizing by class, while the less efficient of the two approaches tried, scaled acceptably well to 16 CPU cores.  Our result indicates that distributed computing frameworks are a viable method for speeding up mutation analysis. 

On research question 2, we observed some connection between  per-module metrics, and the time taken to perform mutation analysis on that module.  Class size has a linear 
relationship with the time taken to conduct mutation analysis on that class; however, that linear relationship is not as strong as one might hope.  By contrast, the proportion of the work required by different mutation operators varied substantially from project to project, and does not appear to be a useful basis for dividing up work efficiently.  

One potential approach given the limited predictive power of these metrics is to use a fine-grained partitioning strategy that produces many more subtasks than  nodes, as Spark will automatically allocate partitions to idle nodes.  Unfortunately, it seems that the overhead for task startup is high, as shown by the relative performance of the two distribution strategies.  The larger partitions from the Parallelize by Mutation Operator strategy were more efficient on small clusters than the Parallelize by Class strategy, though it could not scale as well.  Some of this is due to the unnecessary overheads in our prototype distributed mutation engine.  However, it seems that task startup in Apache Spark is inherently slow, suggesting that the number of partitions should be minimised.

Moreover, our results indicate an alternative work distribution scheme may be feasible.  Mutant generation is extremely fast compared to mutant execution.  As such, generating mutants in the master node, and creating partitions that contain an equal number of mutants in each, means that an appropriate number of work packets can be created to match the cluster size.  The subtasks are likely to require similar amounts of time to execute, thus making very efficient use of the worker nodes.  Mutants can be very compactly represented (in principle, as a byte offset and a mutation operator taking no more than a few 
bytes), and as network bandwidth does not appear to be the limiting factor in performance, sending lists of mutants to the worker nodes is not likely to slow the computation down.  Such an approach will also make it easier to estimate the computation required to conduct mutation analysis, allowing the user to make an informed choice as to whether the time and financial costs of running the analysis are worthwhile.

While parallelization provides a way to overcome one barrier to applying mutation analysis---its slowness---to some extent it may trade the costs of excessive time for the financial costs of renting sufficient cloud infrastructure.  However, as noted in \S\ref{sec:related-work}, most of the existing methodologies for reducing the overhead of mutation analysis can be easily combined with our methods.  Furthermore, the context in which test quality often needs to be assessed---an evolving test suite run regularly on an evolving codebase---offers largely unexplored scope to further reduce the computation required. To our knowledge, no peer-reviewed academic work has examined the optimizations for repeated application of mutation analysis to an evolving codebase and test suite. 
Coles~\cite{coles_pit_2015-2} has suggested several such optimizations for \emph{Incremental Analysis}, of which some have already been implemented in PiT. However, no evidence of the effects of these optimizations on performance and accuracy of the analysis (several of the the proposed optimizations make several possibly unverified assumptions about the behaviour of modified software) has been reported.  We believe that the potential efficiency gains from incremental analysis, including \emph{both}
optimizations already proposed by Coles, and others not yet identified, may be very substantial.  For instance, is seems plausible that a previously killing test will
continue to kill the corresponding mutant the next time mutation analysis is performed; therefore, for previously killed mutants, in most cases only a single tests will need to be executed to verify that the mutant is still killed, instead of executing a significant fraction of all the covering tests.  We plan empirical studies to examine the effect of such optimizations.

Furthermore, in the context of continuous integration, the users of mutation analysis may well be most interested in mutation analysis relating to the parts of the codebase they have changed most recently, rather than the entire codebase.  Therefore, it may prove unnecessary to conduct a complete mutation analysis on each continuous integration step; a ``first pass'' examining mutation coverage for classes that have changed (and possibly their direct dependencies) may be useful, and is likely to be much quicker than  mutation analysis of the entire codebase.  Evaluating
the usefulness of this approach is largely a human factors question, and will require putting a working tool in the hands of developers.

\section{Related Work}
\label{sec:related-work}
Literature on parallelization of mutation analysis dates back to the early 1990s. Offutt \etal~\cite{offutt_mutation_1992} presented a parallel implementation of Mothra on an MIMD system with 16 processing cores. They observed that as, for most systems,
      the number of mutants generated far exceeds the number of test cases, parallelizing by mutant is likely to be a more effective
      approach.  After accounting for the overhead in scheduling and program compilation, they reported an ``almost
      linear'' speed-up.  Similarly promising results were also reported by Choi~\cite{choi_high-performance_1993}.  Modern mutation analysis tools, including PiT~\cite{coles_pit_2015}, often support local parallelization. 
      Source code inspection suggests that PiT parallelizes by mutant.

The only distributed mutation analysis research the authors were able to find was conducted in 1993 by Zapf~\cite{zapf:medusamothra}.  Zapf's MedusaMothra system distributed mutation analysis tasks across a local network of Unix workstations, and sometimes achieved near-linear speedups.  While this farsighted work demonstrated the potential of distributed mutation analysis, the technical context has changed a great deal since.  MedusaMothra partitioned subtasks as finely as is possible - a subtask consisted of a \emph{single} test case and mutant.  Such an approach is likely to be extremely inefficient in a modern context.  Modern unit test suites are written to execute quickly, and individual tests often execute in less than a millisecond.  If Zapf's partitioning approach were tried today, the worker nodes would be likely to spend the vast majority of time idle due to network latency.  

There is a very large research literature on improving the efficiency of serial mutation analysis.  Jia and Harman's survey paper~\cite{jia_analysis_2011} defines two broad categories of approaches: \emph{mutant reduction}, which reduce the number of mutants which must be executed to calculate a mutation score, and \emph{execution reduction}, which reduce the cost of executing a set of mutants.  By this definition both local and distributed parallelism are execution reduction techniques.  Virtually all work in mutant reduction, and much of the serial execution reduction literature, is applicable in a distributed context. 

While there is a rich history of research in mutation analysis, relatively few mutation analysis tools are actively maintained.  PiT, while actively maintained, is less than ideal as a basis for academic research.  Major~\cite{Just2014} is an alternative
mutation analysis system which may be more suitable.  Major is described by its author as ``designed to be highly configurable to support fundamental research in software engineering'', and  ``due to its efficiency and
flexibility, the Major mutation framework is suitable for the application of mutation analysis in research and practice''.  While we did not set out to perform a direct comparison, our results indicate that PiT, even used serially, may still be significantly faster than Major on the same benchmarks.  In part, this may be due to the lack of even local parallelization in the current version.  Regardless, Major's more modular design and research focus may make it a better basis for developing a more advanced distributed mutation analysis system.  

There has been some interest in applying distributed computing frameworks to other testing tasks.  Parveen \etal~\cite{parveen_towards_2009} used Apache Hadoop to speed up execution of JUnit unit test suites.  They reported a thirty-fold speedup over serial execution using a 150-node computing cluster.  Unit testing is a simpler problem than mutation analysis, as the workloads are more predictable.

\section{Conclusion}

We have demonstrated a proof-of-concept tool for performing mutation analysis on distributed cloud-based systems, using Apache Spark and the MapReduce paradigm to distribute the mutation analysis tasks across a cluster of 
Amazon EC2 computing nodes.  Despite known inefficiencies, the prototype was able to outperform the non-distributed tool it was based on on clusters of 8 nodes.  Analysis of the runtime behaviour of our prototype suggests
that while there is some correlation between the size of a module (class) within a codebase and the time required to perform analysis of the mutations in that class, the variation remains substantial.  The minimal cost of 
mutant generation suggests that, instead, the most efficient strategy for distributing work between nodes may be to generate all mutants on the ``master'' node and allocate equal-sized collections of mutants to each worker
node.

Distributed computation is one way in which the long-standing performance barrier to industrial adoption of mutation analysis can be overcome, and the present work, embryonic as it is, demonstrates its potential.  Incremental analysis, 
in the context of a continuous integration workflow, may represent another.  We hope that combining these two techniques with the large existing body of work on speeding up mutation analysis by other means may finally allow it to gain wide industrial adoption.




\bibliographystyle{abbrvurl}
\bibliography{cloud-based-mutation-analysis}
\section*{Supplementary Results}
  \label{apx:supplementary-results}

  This appendix records additional duration results. We only record three trials for each experiment
  due to financial limitations, but the variance between samples appears to be small. Durations for
  the Parallelize by Mutation Operator and Parallelize by Class strategies with 16 processing nodes are shown in Tables
  \ref{tab:supplementary-results:BcBtMm} and \ref{tab:supplementary-results:McBtBm} respectively,
  and durations for unmodified PiT running in parallel with two threads is shown in 
  Table~\ref{tab:supplementary-results:unmodified-pit}.

  \begin{table*}[h]
    \centering
    \begin{tabular}{l|ccc|cc}
    \hline
    & \multicolumn{3}{c|}{\textbf{Trial}} & & \\
    \cline{2-4}
    \textbf{Project} & 1 & 2 & 3 & \textbf{Mean} & \textbf{Standard Deviation} \\
    \hline
    gson & 219561 & 199798 & 220316 & 213225.0 & 11634.0 \\
    joda-time & 737997 & 720419 & 735444 & 731287.0 & 9497.8 \\
    commons-math & 434117 & 433479 & 446328 & 437975.0 & 7241.2 \\
    \hline
    \end{tabular}
    \caption[Durations (Parallelize By Mutation Operator)]{Durations (in milliseconds) for the Parallelize by Mutation Operator strategy on 16
      processing nodes.}
    \label{tab:supplementary-results:BcBtMm}
  \end{table*}

  \begin{table*}[h]
    \centering
    \begin{tabular}{l|ccc|cc}
    \hline
    & \multicolumn{3}{c|}{\textbf{Trial}} & & \\
    \cline{2-4}
    \textbf{Project} & 1 & 2 & 3 & \textbf{Mean} & \textbf{Standard Deviation} \\
    \hline
    gson & 198710 & 215928 & 198961 & 204533.0 & 9869.2 \\
    joda-time & 634838 & 619086 & 630645 & 628190.0 & 8158.0 \\
    commons-math & 252129 & 249744 & 259585 & 253819.0 & 5133.6 \\
    \hline
    \end{tabular}
    \caption[Durations (Parallelize by Class)]{Durations (in milliseconds) for the Parallelize by Class strategy on 16
      processing nodes.}
    \label{tab:supplementary-results:McBtBm}
  \end{table*}

  \begin{table*}[h]
    \centering
    \begin{tabular}{l|ccc|cc}
    \hline
    & \multicolumn{3}{c|}{\textbf{Trial}} & & \\
    \cline{2-4}
    \textbf{Project} & 1 & 2 & 3 & \textbf{Mean} & \textbf{Standard Deviation} \\
    \hline
    gson & 332000 & 344000 & 351000 & 342333.0 & 9609.0 \\
    joda-time & 1081200 & 1087000 & 1081300 & 1083167.0 & 3320.1 \\
    commons-math & 513000 & 514000 & 512000 & 513000.0 & 1000.0 \\
    \hline
    \end{tabular}
    \caption[Durations (Unmodified Parallel Pit)]{Durations (in milliseconds) for unmodified PiT
      running in parallel with two concurrent threads.}
    \label{tab:supplementary-results:unmodified-pit}
  \end{table*}

\balancecolumns 
\end{document}